Occipital and left temporal instantaneous amplitude and frequency oscillations
correlated with access and phenomenal consciousness

Vitor Manuel Dinis Pereira
Language, Mind and Cognition Research Group (LanCog).
Philosophy Centre (CFUL).
Faculty of Letters, University of Lisbon
Alameda da Universidade, 1600-214 Lisboa, Portugal
vpereira1@campus.ul.pt

Abstract

Given the hard problem of consciousness (Chalmers, 1995) there are no
brain electrophysiological correlates of the subjective experience (the felt quality
of redness or the redness of red, the experience of dark and light, the quality of
depth in a visual field, the sound of a clarinet, the smell of mothball, bodily sensations from pains to orgasms, mental images that are conjured up internally, the felt quality of emotion, the experience of a stream of conscious thought or the phenomenology of thought).

However, there are brain occipital and left temporal electrophysiological correlates of the subjective experience (Pereira, 2015).

Notwithstanding, as evoked signal, the change in event-related brain potentials phase (frequency is the change in phase over time) is instantaneous, that is, the frequency will transiently be infinite: a transient peak in frequency (positive or negative), if any, is instantaneous in electroencephalogram averaging or filtering that the event-related brain potentials required and the underlying structure of the event-related brain potentials in the frequency domain cannot be accounted, for example, by the Wavelet Transform (WT) or the Fast Fourier Transform (FFT) analysis, because they require that frequency is derived by convolution rather than by differentiation.

However, as I show in the current original research report, one suitable method for analyse the instantaneous change in event-related brain potentials phase and accounted for a transient peak in frequency (positive or negative), if any, in the underlying structure of the event-related brain potentials is the Empirical Mode Decomposition with post processing (Xie et al., 2014) Ensemble Empirical Mode Decomposition (postEEMD) and Hilbert-Huang Transform (HHT).

Acknowledgements

My mother, Maria Dulce.

Carlos Simões; Diogo Branco; João Carneiro; João Corujo; Manuel Costa;

Manuel Emídio; Sofia Khan; Susana Lourenço.

Language, Mind and Cognition Research Group (LanCog), Philosophy Centre

(CFUL), University of Lisbon: Ricardo Santos.

Introduction

The relevant computation to the effect of the occipital and left temporal correlates of the distinction between access and phenomenology (Pereira, 2015) is the computation of the high degree of visibility "4" and "5" assigned by the participants in both experiments to the correctly identified stimuli (and what there are more in the second experiment is more incorrect answers than in the first experiment), because to distinguish electrophysiologically the access from phenomenology we need that access and phenomenology will be cognitively consciousness of something and we need that access will be the same for all participants in the two experiments (Pereira, 2015: 337-339).

To distinguish electrophysiologically the access from phenomenology we need that access and phenomenology will be cognitively consciousness of something because, for instance, pains are not intentional mental states in the same sense in which cognitive mental states as beliefs and doubts are intentional: there is nothing in sensation other than being felt. That we have a pain and then get there and look at it to talk about it is already our cognition to work, but this (cognitive) way of proceeding does not make the sensation itself an intentional mental state in the same way as when we think about her. Essential to cognitive mental states is that they are intentional without that mark, that of intentionality, they wouldn't be the mental states that in fact they are. But pain is another thing:

what is essential to them is how we feel them, it is their phenomenology - for without it, without phenomenology, they wouldn't be the mental states that in fact they are. And at all, mental states as pains are not about anything, they are not intentional - they don't represent nothing: although we may represent them cognitively in some way, this is not them to be about something, this is us talking about them: this is not them to be cognitions, this is them to be sensations.

That is, we compute the results only from those trials that are the same in the second block of the two experiences.

Saying that phenomenology is not reportable is another way to define phenomenology distinctly from access consciousness, but that we already knew: although confusedly when, for example, it is alleged on the basis of the non reportability the non measurability of phenomenology and thus allegedly that there are not electrophysiological correlates of phenomenology different from the access.

What we know now is what are these electrophysiological correlates of phenomenology different from the access (Pereira, 2015: 344-350): the electrophysiologically correlated with the difference (statistically significant) among those trials that are the same in the second block in both experiments for the same high degree of visibility "4" and "5" (they also access).

The distinct electrophysiological signal correlated with those trials that are the same in the second block in both experiments is phenomenology to be measured (that is, is non reportability to be measured): despite the same high degree of visibility "4" and "5" (as despite the same correct answers).

The high degrees of visibility are the same "4" and "5" but, for those trials that are the same in the second block in both experiments, the statistical difference between the two experiments is significant: because the phenomenology was measured distinctively from the access (behaviorally, there is more incorrect answers in the first trial of the first block of the second experiment, which is distinct from the first trial of the first block of the first experiment in which the masks are to be correctly identified but not the targets as in the case of the first block of the second experiment. The experimental design only differ in the first trial of the first block).

Between 300 and 800 ms, the null hypothesis of latency variability between experiments I and II is not rejected: statistically, the variability of the latency did not significantly differ between the experiments I and II. However the null hypothesis of variability occipital (Oz) and left temporal (T5) amplitude between experiments I and II is rejected, except for the amplitude variability right temporal (T6) between experiments I and II: statistically, the variability of Oz amplitude and T5 differ significantly between experiments I and II, although the variability of amplitude T6 between the experiences I and II does not differ significantly (the null hypothesis of the amplitude variability T6 is accepted).

The second trial of the first block and the second block - in which the stimuli are presented in isolation - are the same in both experiments I and II, thus it is expected that the discrimination of stimulus not statistically significant contrast in correct and incorrect responses between the two experiments (Pereira, 2015).

In isolated presentations of targets, there is no statistically significant contrast in stimuli discrimination between the two experiment II [$\chi(1)$ = 3.492, p = .062, 0 cells (.0%) have expected count less than 5. The minimum expected count is 17.47] and experiment I [$\chi(1)$ = 3.160, p = .075, 0 cells (.0%) have expected count less than 5. The minimum expected count is 20.38], the contrast in stimuli discrimination between more correct responses in experiment II (square correct 97.1%, incorrect 2.9 %; diamond correct 98.5%, incorrect 1.5 %) than in experiment I (square correct 96.7%, incorrect 3.3 %; diamond correct 98.1 %, incorrect 1.9 %) is not statistically significant, and so, there are not a difference statistically significant between correct responses in experiment II than in experiment I (the block in which the targets are presented in isolation is the same second in both experiments I and II). The access not significantly change between experiment II and experiment I, the access remains the same. (Pereira, 2015.)

The difference between the two experiments in mean rank within the interval of degrees of visibility "4" and "5" remains statistically significant between the two experiments in targets isolated presentations correctly identified [H(1) = 336.045, p = 0.000), with a mean rank of 1081.05 for experiment I, and a mean rank of 1522.38 for experiment II (Kruskal-Wallis Test)]. (Pereira, 2015.)

Targets isolated presentations correctly identified, because to distinguish electrophysiologically the access from phenomenology we need that access and phenomenology will be cognitively consciousness of something and we need that access will be the same for all participants in the two experiment.

Notwithstanding, as evoked signal, the change in event-related brain potentials (ERPs) phase (frequency is the change in phase over time) is instantaneous, that is, the frequency will transiently be infinite: a transient peak in frequency (positive or negative), if any, is instantaneous in electroencephalogram (EEG) averaging or filtering that the ERPs required and the underlying structure of the ERPs in the frequency domain cannot be accounted, for example, by the Wavelet Transform (WT) or the Fast Fourier Transform (FFT) analysis, because they require that frequency is derived by convolution (frequency are pre-defined and constant over time) rather than by differentiation (without predefining frequency and accounted that frequency may vary over time).

Participants, Apparatus and stimuli, Procedure, EEG recording, Behavioral data

As in Pereira (2015).

Twenty two adults with normal vision or corrected to normal, without neurological or        psychiatric history, ignoring completely the experimental purposes.

Five participants were excluded dues to excessive EEG artifacts (3) or insufficient trials       (2).

The experimental protocol was approved by the doctoral program in Cognitive Science,      University of Lisbon.

Two types of targets: square (1.98 cm side) or diamond (for 45 ° rotation of the square).

Two types of masks: mask or pseudo-mask.

The width of the mask is 3.05 cm and its inner white portion (RGB 255-255-255) is 8       mm wider than the black (RGB 0-0-0) target stimulus.

The width of the pseudo-mask is 3.10 cm and its inner white portion is circular (2.63 cm diameter).

Despite the different sizes, the color black stands in the same area, both in mask and pseudo-mask, and its luminance is identical. This was expected to be important to make the masks produce similar ERPs when presented alone.

All stimuli are presented on a gray background (RGB 173-175-178).

First task: to recognize which of the targets – square or diamond – is presented.

Second task: to evaluate the visibility of targets.

The answers are given using the keyboard or the mouse.

In the first task, recognition of targets, participants respond if "they seemed to have seen something" by pressing "8" to "yes" and "9" to "no". A negative response completes the trial and starts the next. An affirmative answer conduces the subjects to a screen of sixteen stimuli to signal with the mouse what seems they have been identified. The position on the screen indicated by the participant will be recorded informatically as coordinate system <X,Y>.

In the second task (evaluation of visibility) we used a Likert scale from "1" to "5": "not visible at all" ("1" key), "barely visible" ("2" key), "visible, but obscure" (key "3 "), "clear but not quite visible" (key "4 ") and "perfectly clear and visible" ("5" key).

Experiments were held at the Faculty of Psychology in a slightly darkened silent room. Participants were seated in a reclining chair at 81.28 centimeters distance from the 50.8 centimeters monitor.

It is expected that the running of experiment train the volunteer. The beginning of the behavioral and EEG recording is unknown to the volunteer.

The SuperLab program for Windows from Cedrus, PC - compatible, connected to a       SVGA color monitor, manages the presentation of stimuli, randomizes their sequence (the trials in each block), the exposure times, the record of the response, triggers the       trigger synchronize with the system acquisition of physiological signals, MP100 and   EEG       amplifiers,       program AcqKnowledge, both of Biopac.

Experiment I.

Eight volunteers (aged 18–46 years, M= 22.50, SD=9.562, 7 females).

The target and the masks will be presented for 17 ms. The mask (or pseudo-mask) appears 1 ms after the presentation of the target (inter-stimulus interval, ISI, the interval       between the termination of the target and the onset of the mask). These ISI (1 ms)       correspond to 18 ms stimulus-onset asynchrony (SOA, the interval between the onset of   the target and the mask) (rounded values). Answers were signaled by mouse on a screen       of sixteen stimuli, among which are the mask and pseudo-mask. Note that the subject       not performed a forced-choice task, for example, reading any question either ''Diamond       or Square?'' or ''Square or Diamond?'', even if counterbalanced across participants (contrast with, for example, Lau and Passingham 2006).

In the second trial, masks were presented for 17 ms, and answers were signaled by   mouse on a screen of sixteen stimuli, among which are the mask and pseudo-mask.

Second block. Trial: targets will be presented for 17 ms, and answers were signaled by   mouse on screen of sixteen stimuli, among which are the targets.

Experiment II.

Nine volunteers (aged 20–26 years, M= 21.22, SD=2.224, 5 females).

The target will be presented for 17 ms (like experiment I), but the mask (or pseudo-          mask) will be presented for 167 ms (unlike experiment I). Note that in all experiments,     targets are always shown for 17 ms and is never replaced, for example, by a blank  screen  with  the  same  duration  of  17  ms  (contrast,  for example, with Del Cul et al. 2007).

Unlike experiment I, the target is intercalated between two presentations of the      mask/pseudo-mask (each, 167 ms): one earlier, paracontrast; the other after target,    metacontrast.

The mask (or pseudo-mask) appears 0 ms before (forward masking) and 1 ms after        (backward  masking)  the  presentation  of  the  target  (inter-stimulus interval, ISI, the        interval between the termination of the target and the onset of the mask). These ISI (1     ms) correspond to 18 ms stimulus-onset asynchrony (SOA, the interval between the        onset of the target and the mask) and to 168 ms stimulus-termination asynchrony (STA,     the interval between the termination of the target and of the mask) (rounded values).      Unlike  experiment  I,  the  answer were signaled by mouse on screen of sixteen stimuli,      among   which   are   the targets.

In the second trial, like experiment I: masks were presented for 17 ms, and answers       were signaled by mouse on a screen of sixteen stimuli, among which are the mask and      pseudo-mask.

Second block, like experiment I. Trial: targets will be presented for 17 ms, and answers  were signaled by mouse on screen of sixteen stimuli, among which are the targets.

The Empirical Mode Decomposition (EMD) with post processing (Xie et al. 2014)

and Hilbert-Huang Transform (HHT)

Now, we move from the features of the ERP (such as the amplitude and latency of peaks) of the Pereira (2015) towards the decomposition of amplitude and to the instantaneous frequency resulting from this decomposition: the instantaneous amplitude and the instantaneous frequency of event-related changes correlated with a contrast in access and with a contrast in phenomenology.

Despite that the Wavelet or the Fourier Transform are the methods most widely used for analysing the linear (proportionality or additivity) and stationary (the signal, and so the time series representing this signal, has the same mean and variance throughout) properties of the EEG signal, the EEG signal have nonlinear (nonproportionality or nonadditivity) and non stationary (signal's statistical characteristics change with time) properties.

However, one suitable method for analyse the instantaneous change in event-related brain potentials (ERPs) phase and accounted for a transient peak in frequency (positive or negative), if any, in the underlying structure of the ERPs is the Empirical Mode Decomposition (EMD) with post processing (Xie et al., 2014) Ensemble Empirical Mode Decomposition (postEEMD) and Hilbert-Huang Transform (HHT).

The Wavelet or the Fourier Transform analyse the signal in time-frequency-energy (Wavelet) and in frequency-energy (Fourier) domain without discrete feature extraction (Wavelet, with continuous feature extraction) or without discrete or continuous feature extraction (Fourier).

However, the Hilbert-Huang Transform (HHT) analyse the signal in time-

frequency-energy domain for feature extraction.

For example, either the Fourier functions or the EMD functions are oscillations with zero mean derived from the decomposition of a signal (for example, ERPs) that when summed together reconstitute the original signal.

However, whereas the Fourier functions are called harmonic functions meaning that they amplitude and frequency are constant over time, the EMD functions are called Intrinsic Mode Functions (IMFs) meaning that they amplitude and frequency may vary over time.

Once the Intrinsic Mode Functions have been extracted and post processing (Xie et al., 2014), the Hilbert-Huang Transform can be used to display the underlying structure in the amplitude and frequency domain of the grand average occipital and left temporal electrical activity characterized in Pereira (2015).

These insights may prove to be a useful a guide in helping us move from a focus on the features of the ERP, such as the amplitude and latency of peaks (Pereira 2015), towards a study of the amplitude, instantaneous frequency and energy structure of event-related changes over time in the EEG.

Given the EEG recording in Pereira (2015) with a duration of 1150 ms, defined as 150 ms before the stimulus (baseline) and 1000 ms after its occurrence, there are 16 ERPs correlated with combined target-mask presentations and with isolated presentations of square or diamond and of mask or pseudo-mask: 8 by each channels, occipital and left temporal.

The Empirical Mode Decomposition (EMD) with post processing (Xie et al., 2014) and  the Hilbert-Huang Transform resulted in 56 occipital and 56 left temporal variables (7 postIMFs by each of the 8 ERPs) in amplitude domain and resulted in 56 occipital and 56 left temporal variables in frequency domain with 460 observations each: 230 observations (the EEG records duration of 1150 ms) by each of two experiments. In the energy domain, the resulted is 8 occipital and 8 left temporal variables (ERPs) with 14 observations each (7 postIMFs by each of two experiments).

## Calculated the variance inflation factor (VIF)

To evaluate the presumed excessive correlation among variables (i.e., colinearity), we calculated the variance inflation factor (VIF) for each variable by vif_fun.r (beckmw 2013, February 5).

If the VIF calculated for each variable is more than 10 (values in the range of 5-10 are commonly used as threshold), colinearity is strongly suggested, and the variable removed.

This calculations reduce the 56 variables to 32 in occipital amplitude domain, reduce the 56 variables to 28 in left temporal amplitude domain, reduce the 56 variables to 51 in occipital frequency domain and reduce the 56 variables to 51 in left temporal frequency domain. In energy domain, the calculated variance inflation factor (VIF) removed 4 variables excessive correlated.

Figures 1-2 show the underlying structure in the amplitude domain of the grand average occipital and left temporal electrical activity correlated with a contrast in access (characterized in Pereira 2015: 340-344), after the calculated variance inflation factor (VIF) removed respectively 11 (Oz combined target-mask presentations) and 14 (T5 combined target-mask presentations) variables excessive correlated. The 11 Oz combined target-mask variables excessive correlated are postIMF 4 (SquarePseudo,DiamondPseudo), postIMF5 (SquareMask,

SquarePseudo), postIMF 6 (SquareMask, DiamondMask, DiamondPseudo) and postIMF 7 (SquareMask, SquarePseudo, DiamondMask and DiamondPseudo). The 14 T5 combined target-mask variables excessive correlated are postIMF 3 (SquareMask, DiamondMask), postIMF 4 (SquereMask, DiamondMask,DiamondPseudo), postIMF5 (SquareMask, SquarePseudo, DiamondMask, DiamondPseudo), postIMF 6 (SquareMask, SquarePseudo, DiamondMask, DiamondPseudo) and postIMF 7 (DiamondMask ).

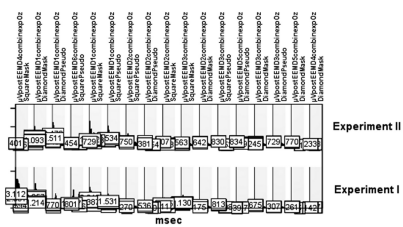

Fig. 1. Related to combined target-mask presentations, the underlying hilbert structure in the amplitude domain of the grand average occipital (Oz) electrical activity correlated with a contrast in access (characterized in Pereira 2015: 340-344), after the calculated variance inflation factor (VIF) removed 11 variables excessive correlated.

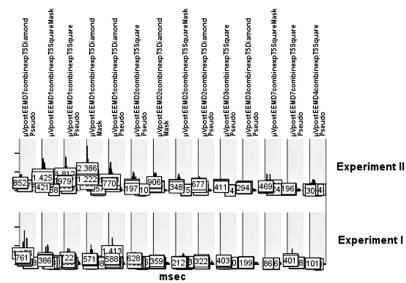

Fig. 2. Related to combined target-mask presentations, the underlying hilbert structure in the amplitude domain of the grand average left temporal (T5) electrical activity correlated with a contrast in access (characterized in Pereira 2015: 340-344), after the calculated variance inflation factor (VIF) removed 14 variables excessive correlated.

Figures 3-6 show the underlying structure in the amplitude domain of the grand average occipital and left temporal electrical activity correlated with a contrast in phenomenology (characterized in Pereira 2015: 344-350), after the calculated variance inflation factor (VIF) removed respectively 4 (Ozsqua), 4 (Ozdiamo), 3 (T5squa) and 5 (T5diamo) variables excessive correlated. The 4 Ozsqua isolated target variable excessive correlated are postIMF 4, 5, 6 and 7. The 4 Ozdiamo isolated target variable excessive correlated are postIMF 3, 4, 5 and 6. The 3 T5squa isolated target variable excessive correlated are postIMF 4, 5 and 7. The 5 T5diamo isolated target variable excessive correlated are postIMF 3, 4, 5, 6 and 7.

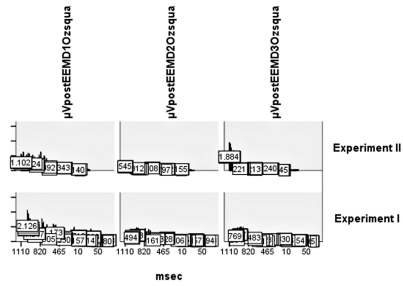

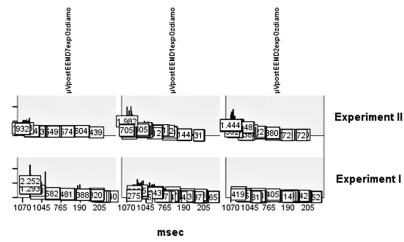

Fig. 3. Related to isolated square presentations, the underlying hilbert structure in the amplitude domain of the grand average occipital (Oz) electrical activity correlated with a contrast in phenomenology (characterized in Pereira 2015: 344-360), after the calculated variance inflation factor (VIF) removed 4 variables excessive correlated.

Fig. 4. Related to isolated diamond presentations, the underlying hilbert structure in the amplitude domain of the grand average occipital (Oz) electrical activity correlated with a contrast in phenomenology (characterized in Pereira 2015: 344-360), after the calculated variance inflation factor (VIF) removed 4 variables excessive correlated.

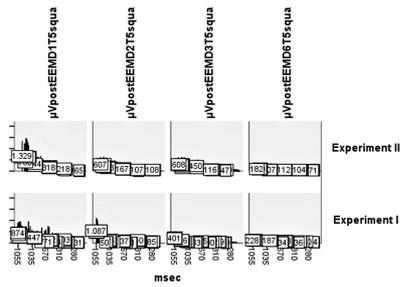

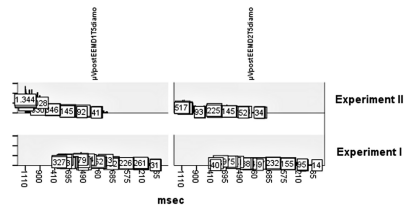

Fig. 5. Related to isolated square presentations, the underlying hilbert structure in the amplitude domain of the grand average left temporal (T5) electrical activity correlated with a contrast in phenomenology (characterized in Pereira 2015: 344-360), after the calculated variance inflation factor (VIF) removed 3 variables excessive correlated.

Fig. 6. Related to isolated diamond presentations, the underlying hilbert structure in the amplitude domain of the grand average left temporal (T5) electrical activity correlated with a contrast in phenomenology (characterized in Pereira 2015: 344-360), after the calculated variance inflation factor (VIF) removed 5 variables excessive correlated.

Figures 7-8 show the underlying structure in the frequency domain of the grand average occipital and left temporal electrical activity correlated with a contrast in access (characterized in Pereira 2015: 340-344), after the calculated variance inflation factor (VIF) removed respectively 4 (Oz combined target-mask presentations) and 4 (T5 combined target-mask presentations) variables excessive correlated. The 4 Oz combined target-mask variables excessive correlated are postIMF 7 (SquareMask, SquareMask, DiamondMask and DiamondPseudo). The 4 T5 combined target-mask variables excessive correlated are postIMF 6 (SquareMask), postIMF 7 (SquareMask), postIMF 7 (DiamondMask and DiamondPseudo).

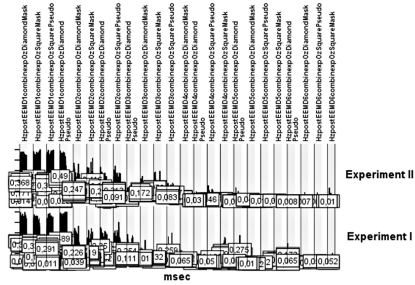 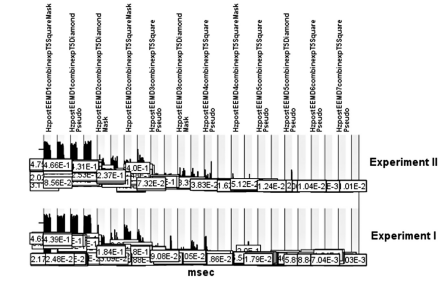

Fig. 7. Related to combined target-mask presentations, the underlying hilbert structure in the frequency domain of the grand average occipital (Oz) electrical activity correlated with a contrast in access (characterized in Pereira 2015: 340-344), after the calculated variance inflation factor (VIF) removed 4 variables excessive correlated.

Fig. 8. Related to combined target-mask presentations, the underlying hilbert structure in the frequency domain of the grand average left temporal (T5) electrical activity correlated with a contrast in access (characterized in Pereira 2015: 340-344), after the calculated variance inflation factor (VIF) removed 4 variables excessive correlated.

Figures 9-12 show the underlying structure in the frequency domain of the grand average occipital and left temporal electrical activity correlated with a contrast in phenomenology (characterized in Pereira 2015: 344-350), after the calculated variance inflation factor (VIF) removed 1 (postIMF 6 T5squa) variables excessive correlated: any OzSquare, Ozdiamo or T5diamo variables is removed because neither variable is excessive correlated.

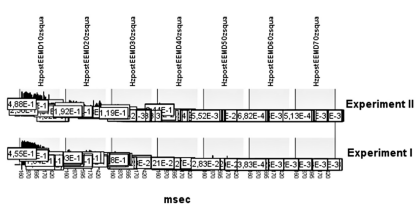 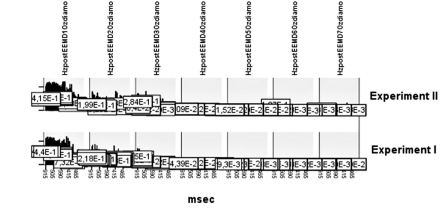

Fig. 9. Related to isolated square presentations, the underlying hilbert structure in the frequency domain of the grand average occipital (Oz) electrical activity correlated with a contrast in phenomenology (characterized in Pereira 2015: 344-350), after the calculated variance inflation factor (VIF): any variable is removed because neither variable is excessive correlated.

Fig. 10. Related to isolated diamond presentations, the underlying hilbert structure in the frequency domain of the grand average occipital (Oz) electrical activity correlated with a contrast in phenomenology (characterized in Pereira 2015: 344-350), after the calculated variance inflation factor (VIF): any variable is removed because neither variable is excessive correlated.

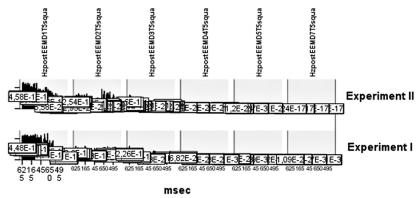 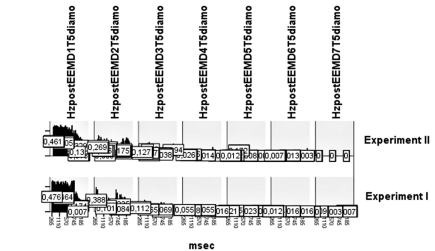

Fig. 11. Related to isolated square presentations, the underlying hilbert structure in the frequency domain of the grand average left temporal (T5) electrical activity correlated with a contrast in phenomenology (characterized in Pereira 2015: 344-350), after the calculated variance inflation factor (VIF) removed 1 variable excessive correlated.

Fig. 12. Related to isolated diamond presentations, the underlying hilbert structure in the frequency domain of the grand average left temporal (T5) electrical activity correlated with a contrast in phenomenology (characterized in Pereira 2015: 344-350), after the calculated variance inflation factor (VIF): any variable is removed because neither variable is excessive correlated.

Partial least squares regression (PLSR): the minimal root mean squared error of

prediction (RMSEP)

If we set 59 as the seed, the partial least squares regression (PLSR) (Wold 2001, Martens 2001, Mevik and Wehrens 2007), cross-validated using 10 random segments, returned the postIMF 3 (combinexpOzDiamondPseudo) as the minimal root mean squared error of prediction (RMSEP) for Oz instantaneous amplitude (23 variables) and the postIMF 2 (combinexpT5DiamondPseudo) as the minimal root mean squared error of prediction (RMSEP) for T5 instantaneous amplitude (19 variables).

The minimal postIMF that least erroneously explains the variability between the two experiments in Oz instantaneous amplitude is predictably the postIMF 3 combined Diamond Pseudo. That is, the minimal postIMF instantaneous amplitude associated (with less error of prediction) to the variability between the two experiments in Oz instantaneous amplitude is, after partial least squares regression validation, postIMF 3 combined Diamond Pseudo. The postIMF 3 combined Diamond Pseudo is the Oz instantaneous amplitude minimal value that we can use to measure with less error of prediction the propagation of the remaining Oz instantaneous amplitude values around the variability between the two experiments. (Fig. 13.)

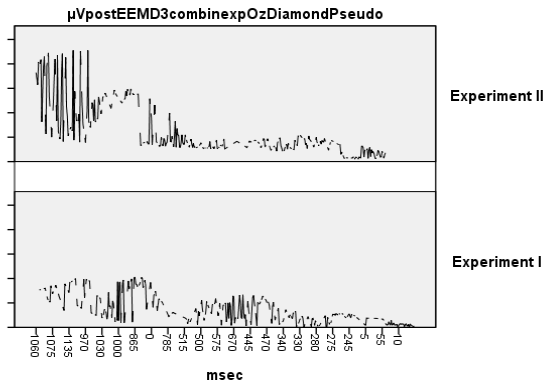

Fig. 13. The postIMF 3 combined Diamond Pseudo is the Oz instantaneous amplitude minimal value
that we can use to measure with less error of prediction the propagation of the remaining Oz
instantaneous amplitude values around the variability between the two experiments.

An independent-sample t-test was conducted to compare postIMF 3 combined Diamond Pseudo (Ozp3cDP) between the two experiments in Oz instantaneous amplitude. Equal variances not assumed, there was a significant difference in the postIMF 3 combined Diamond Pseudo Oz instantaneous amplitude for experiment II (M= 250.08, SD= 221. 24) and experiment I (M=141.85, SD = 116. 99), t (347.796)= 6.559, p < 0.001, 95% CI [75.78, 140.69], g [ 95 % CI] = 0.61 [ 0.42 , 0.8 ]. The Common Language Effect Size (CLES) indicates that the chance that for a randomly selected pair of Ozp3cDP instantaneous amplitude values the Ozp3cDP instantaneous amplitude values from experiment II is higher than the Ozp3cDP instantaneous amplitude values from experiment I is 66.7% (Del Re, 2013).

The minimal postIMF that least erroneously explains the variability between the two experiments in T5 instantaneous amplitude is predictably the postIMF 2 combined Diamond Pseudo. That is, the minimal postIMF instantaneous amplitude associated (with less error of prediction) to the variability between the two experiments in T5 instantaneous amplitude is, after partial least squares regression validation, postIMF 2 combined Diamond Pseudo. The postIMF 2 combined Diamond Pseudo is the T5 instantaneous amplitude minimal value that we can use to measure with less error of prediction the propagation of the remaining T5 instantaneous amplitude values around the variability between the two experiments. (Fig. 14.)

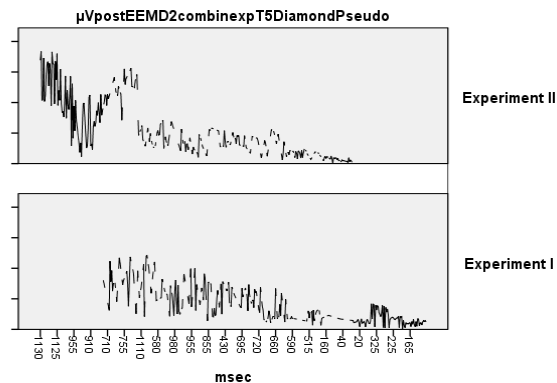

Fig. 14. The postIMF 2 combined Diamond Pseudo is the T5 instantaneous amplitude minimal value that we can use to measure with less error of prediction the propagation of the remaining T5 instantaneous amplitude values around the variability between the two experiments.

An independent-sample t-test was conducted to compare postIMF 2 combined Diamond Pseudo (T5p2cDP) between the two experiments in T5 instantaneous amplitude. Equal variances not assumed, there was a significant difference in the postIMF 2 combined Diamond Pseudo T5 instantaneous amplitude for experiment II (M= 231.50 SD= 193.05) and experiment I (M=148.88, SD = 116.23), t (375,748)= 5,561, p < 0.001, 95% CI [53.40, 111.84], g [ 95 % CI] = 0.52 [ 0.33 , 0.7 ]. The Common Language Effect Size (CLES) indicates that the chance that for a randomly selected pair of T5p2cDP instantaneous amplitude values the T5p2cDP instantaneous amplitude values from experiment II is higher than the T5p2cDP instantaneous amplitude values from experiment I is 64.28% (Del Re, 2013).

If we set 59 as the seed, the partial least squares regression (PLSR), cross-validated using 10 random segments, returned the postEEMD 1 (OzMask) as the minimal root mean squared error of prediction (RMSEP) for Oz instantaneous frequency (40 variables) and the postIMF 5 (T5Mask) as the minimal root mean squared error of prediction (RMSEP) for T5 instantaneous frequency (43 variables).

The minimal postIMF that least erroneously explains the variability between the two experiments in Oz instantaneous frequency is predictably the postIMF 1 Mask. That is, the minimal postIMF instantaneous frequency associated (with less error of prediction) to the variability between the two experiments in Oz instantaneous frequency is, after partial least squares regression validation, postIMF 1 Mask. The postIMF 1 Mask is the Oz instantaneous frequency minimal value that we can use to measure with less error of prediction the propagation of the remaining Oz instantaneous frequency values around the variability between the two experiments. (Fig. 15.)

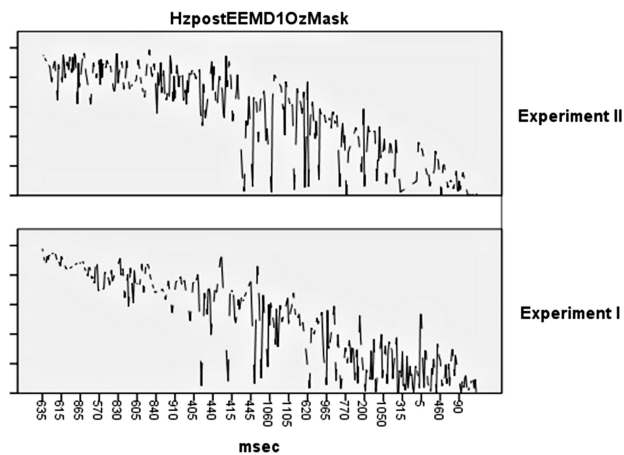

Fig. 15. The postIMF 1 Mask is the Oz instantaneous frequency minimal value that we can use to measure with less error of prediction the propagation of the remaining Oz instantaneous frequency values around the variability between the two experiments.

An independent-sample t-test was conducted to compare postIMF 1 Mask (Ozp1M) between the two experiments in Oz instantaneous frequency. Equal variances not assumed, there was a significant difference in the postIMF 1 Mask Oz instantaneous frequency for experiment II (M= 0.29, SD= 0.14) and experiment I

(M=0.23, SD = 0.15), t (457.061)= 4.737, p < 0.001, 95% CI [0.03, 0.09], g [ 95 % CI] = 0.44 [ 0.26 , 0.63 ]. The Common Language Effect Size (CLES) indicates that the chance that for a randomly selected pair of Ozp1M instantaneous frequency values the Ozp1M instantaneous frequency values from experiment II is higher than the Ozp1M instantaneous frequency values from experiment I is 62.24% (Del Re, 2013).

The minimal postIMF that least erroneously explains the variability between the two experiments in T5 instantaneous frequency is predictably the postIMF 5 Mask. That is, the minimal postIMF instantaneous frequency associated (with less error of prediction) to the variability between the two experiments in T5 instantaneous frequency is, after partial least squares regression validation, postIMF 5 Mask. The postIMF 5 Mask is the T5 instantaneous frequency minimal value that we can use to measure with less error of prediction the propagation of the remaining T5 instantaneous frequency values around the variability between the two experiments. (Fig. 16.)

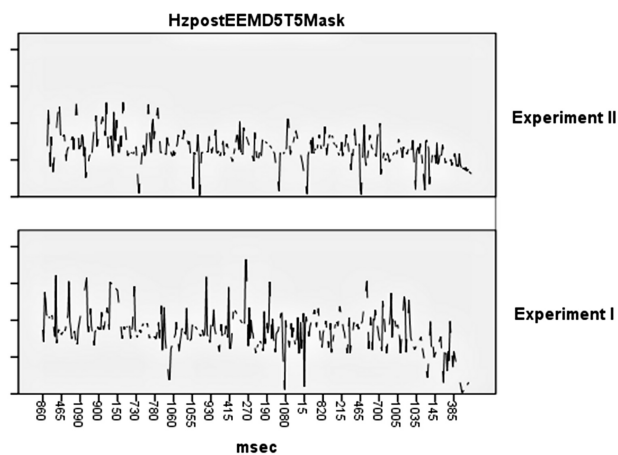

Fig. 16. The postIMF 5 Mask is the T5 instantaneous frequency minimal value that we can use to measure with less error of prediction the propagation of the remaining T5 instantaneous frequency values around the variability between the two experiments.

An independent-sample t-test was conducted to compare postIMF 5 Mask (T5p5M) between the two experiments in T5 instantaneous frequency. Equal variances not assumed, there was a significant difference in the postIMF 5 Mask T5 instantaneous frequency for experiment I (M=0.017, SD = 0.007) and experiment II (M= 0.0129, SD= 0.0046), t (440.573)= 9.011 p < 0.001,  95% CI [ 0.0034,0.0053], g [ 95 % CI] = 0.84 [ 0.65 , 1.03 ]. The Common Language Effect Size (CLES) indicates that the chance that for a randomly selected pair of T5p5M instantaneous frequency values the T5p5M instantaneous frequency values from experiment I is higher than the T5p5M instantaneous frequency values from experiment II is 72.35% (Del Re, 2013).

The calculated minimal value that we can use to measure with less error of prediction the propagation of the remaining values around the variability between the two experiments reduced the number of variables in instantaneous amplitude from 32 (of 56, after variance inflation factor calculations) to 23 (Oz) and from 28 (of 56, after variance inflation factor calculations) to 19 (T5) and reduced the number of variables in instantaneous frequency from 51 (of 56, after variance inflation factor calculations) to 40 (Oz) and from 51 (of 56, after variance inflation factor calculations) to 43 (T5).

Notwithstanding, what variables are important for the variability between the two experiments remained to be assessed.

Partial least squares regression (PLSR): significance multivariate correlation (sMC) statistic

Given the calculated minimal value that we can use to measure with less error of prediction (namely, 23 variables in Oz instantaneous amplitude, 19 variables in T5 instantaneous amplitude, 40 variables in Oz instantaneous frequency and 43 variables in T5 instantaneous frequency) the propagation of the remaining values around the variability between the two experiments, which variables are important for the variability between the two experiments are assessed by significance multivariate correlation (sMC) statistic (Afanador et al., 2016 and, e.g., Thanh et al., 2014) of the partial least squares regression (PLSR) results (figs. 13-16), cross-validated using 10 random segments (setting 59 as the seed).

In other words, which variables are important for the variability between the two experiments are assessed by comparing the ratios between the variable-wise Mean Squared Errors (MSE) and the mean squared of its residuals to an F-test with 1 and N - 2 degrees of freedom (the cut-off is based on the F-test, because appeared that the cut-off based on the mean was influencing negatively the predictions): the variables that exceed the F-test threshold are selected.

If we set 59 as the seed, the significance multivariate correlation (sMC) statistic, with a correction of 1st order auto-correlation in the residuals, "out-of-bag" (OOB) validation and with 1000 cross-validation bootstrap samples, selected the following variables as important for the variability between the two experiments in Oz and T5 instantaneous amplitude and in Oz and T5 instantaneous frequency.

Related to Oz instantaneous amplitude, the 4 variables postIMF 6 SquarePseudo, postIMF 7 diamo, postIMF 4 SquareMask, postIMF 4 DiamondMask, empirical decomposed, post processing (Xie et al., 2014) and Hilbert-Huang transformed from Oz event-related changes (Pereira, 2015), are selected as important for the variability between the two experiments in instantaneous amplitude.

Related to T5 instantaneous amplitude, the 2 variables postIMF 7 DiamondPseudo and postIMF 7 SquarePseudo, empirical decomposed, post processing (Xie et al. 2014) and Hilbert-Huang transformed from T5 event-related changes (Pereira, 2015), are selected as important for the variability between the two experiments in instantaneous amplitude.

Related to Oz instantaneous frequency, the 6 variables postIMF 7 Pseudomask, postIMF 6 SquarePseudo, postIMF 5 SquarePseudo, postIMF 5 SquareMask, postIMF 6 DiamondPseudo and postIMF 7 diamo, empirical decomposed, post processing (Xie et al. 2014) and Hilbert-Huang transformed from Oz event-related changes (Pereira, 2015), are selected as important for the variability between the two experiments in instantaneous frequency.

Related to T5 instantaneous frequency, the 8 variables postIMF 7 squa, postIMF 7 Mask, postIMF 5 SquarePseudo, postIMF 5 diamo, postIMF 5 SquareMask, postIMF 4 diamo, postIMF 4 SquarePseudo and postIMF 5 DiamondMask, empirical decomposed, post processing (Xie et al., 2014) and Hilbert-Huang transformed from T5 event-related changes (Pereira, 2015), are selected as important for the variability between the two experiments in instantaneous frequency.

The repeated measures analysis of variance (ANOVA) with the experiment I and experiment II (Pereira, 2015) as a between-subjects factors and the postIMF variables selected as important by significance multivariate correlation (sMC) statistic as a within-subject factors gave the following significant (Greenhouse-Geisser correction for violations of the sphericity) results for Oz instantaneous amplitude [F(1.197,548.082)= 146.612, p < 0.001, $\eta p^2$ = 0.24249, 90% CI [0.96 , 1.29], $\eta G^2$ = 0.14940] and for T5 instantaneous amplitude [F(1, 458)= 3710.346, p

< 0.001, $\eta p^2$ = 0.89012, 90% CI [5.33 , 6.02], $\eta G^2$ = 0.51299]. For Oz and T5 instantaneous frequency, the repeated measures ANOVA results are respectively [F(3.111,1424.755)= 73.586, p < 0.001, $\eta p^2$ = 0.13843, 90% CI [0.64 , 0.96], $\eta G^2$ = 0.08082] (Oz) and [F(2.493,1141.681)= 441.582, p < 0.001, $\eta p^2$ = 0.49087, 90% CI [1.77 , 2.14], $\eta G^2$ = 0.42934] (T5) (Lakens, 2013).

The statistically significant contrast in the variability of intrinsic mode functions, empirical decomposed and post processing (Xie et al., 2014) from event-related changes (Pereira 2015), between the two experiments correlated with a contrast in access is for the instantaneous amplitude within the 3 variables postIMF 6 SquarePseudo, postIMF 4 SquareMask, postIMF 4 DiamondMask (Oz) (fig. 17) and within the 2 variables postIMF 7 DiamondPseudo and postIMF 7 SquarePseudo (T5) (fig. 18).

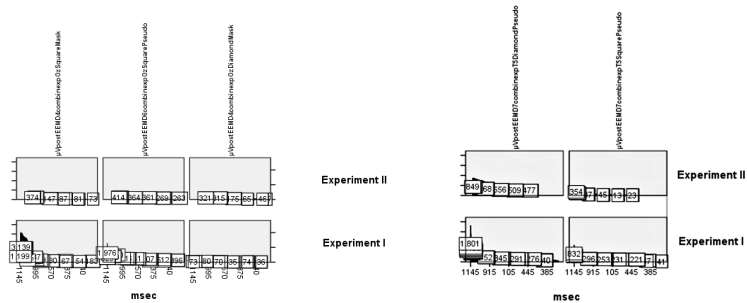

Fig. 17. The statistically significant contrast in the variability of intrinsic mode functions between the two experiments correlated with a contrast in access is for the instantaneous amplitude within the 3 variables postIMF 6 SquarePseudo, postIMF 4 SquareMask, postIMF 4 DiamondMask (Oz).

Fig. 18. The statistically significant contrast in the variability of intrinsic mode functions between the two experiments correlated with a contrast in access is for the instantaneous amplitude within the 2 variables postIMF 7 DiamondPseudo and postIMF 7 SquarePseudo (T5).

Related to instantaneous frequency, the statistically significant contrast in the variability of intrinsic mode functions between the two experiments correlated with a contrast in access is within the 4 variables postIMF 6 SquarePseudo, postIMF 5 SquarePseudo, postIMF 5 SquareMask and postIMF 6 DiamondPseudo (Oz) (fig. 19) and within the 4 variables postIMF 5 SquarePseudo, postIMF 5 SquareMask, postIMF 4 SquarePseudo and postIMF 5 DiamondMask (T5) (fig. 20).

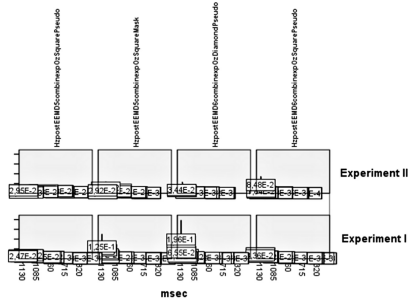 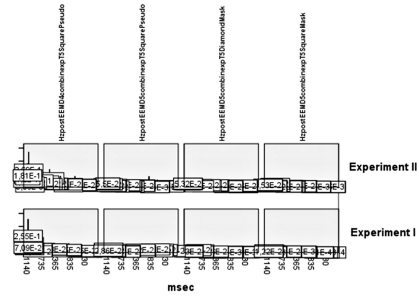

Fig. 19. The statistically significant contrast in the variability of intrinsic mode functions between the two experiments correlated with a contrast in access is for instantaneous frequency within the 4 variables postIMF 6 SquarePseudo, postIMF 5 SquarePseudo, postIMF 5 SquareMask and postIMF 6 DiamondPseudo (Oz).

Fig. 20. The statistically significant contrast in the variability of intrinsic mode functions between the two experiments correlated with a contrast in access is for instantaneous frequency within the 4 variables postIMF 5 SquarePseudo, postIMF 5 SquareMask and postIMF 4 SquarePseudo and postIMF 5 DiamondMask (T5).

The statistically significant variability between the two experiments correlated with a contrast in phenomenology is for the instantaneous amplitude within the 1 variables postIMF 7 diamo (Oz) (fig. 21).

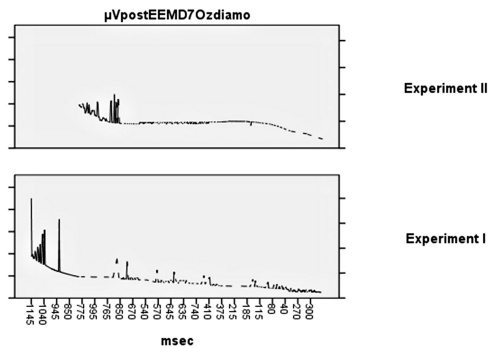

Fig. 21. The statistically significant variability between the two experiments correlated with a contrast in phenomenology is for the instantaneous amplitude within the 1 variables postIMF 7 diamo (Oz).

However, none variable, empirical decomposed, post processing (Xie et al., 2014) and Hilbert-Huang transformed from T5 event-related changes (Pereira, 2015), is within selected as important for the variability in instantaneous amplitude between the two experiments correlated with a contrast in phenomenology, as it is selected as important for the variability in T5 instantaneous frequency (below after the next paragraph, the 2nd next paragraph).

Remind that the trials that are the same in the second block in both experiments for the same high degree of visibility "4" and "5" (they also access) and for the same correct answers (stimulus's discrimination don't contrast in correct and incorrect responses between the two experiments) are the isolated presentations of square or diamond and of mask or pseudo-mask.

Related to instantaneous frequency, the statistically significant contrast in the variability of intrinsic mode functions between the two experiments correlated with a contrast in phenomenology is within the 1 variables postIMF 7 diamo (Oz) (fig. 22) and within the 3 variables postIMF 7 squa, postIMF 5 diamo, postIMF 4 diamo (T5) (fig. 23).

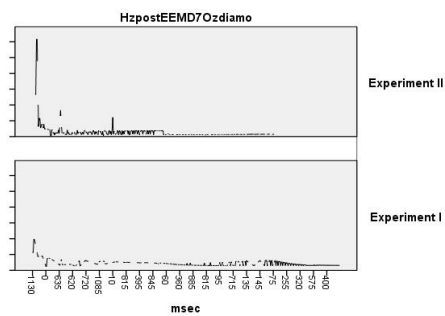 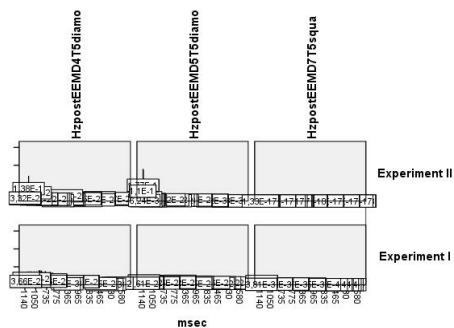

Fig. 22. The statistically significant variability between the two experiments correlated with a contrast in phenomenology is for the instantaneous frequency within the 1 variables postIMF 7 diamo (Oz).

Fig. 23. The statistically significant variability between the two experiments correlated with a contrast in phenomenology is for the instantaneous frequency within the 3 variables postIMF 7 squa , postIMF 5 diamo , postIMF 4 diamo (T5).

These intrinsic mode functions explain the variability of the occipital and left temporal electrical activity co-occurring with a contrast in access distinctly from the variability of the occipital and left temporal electrical activity co-occurring with a contrast in phenomenology (Pereira, 2015)  with  the accuracy that,  related to  the left temporal (T5) electrical activity co-occurring with a contrast in phenomenology, the distinct  electrophysiological signal  is in the instantaneous  frequency domain but don't in T5 instantaneous amplitude domain.